# COVID-19 Detection Using Slices Processing Techniques and a Modified Xception Classifier from Computed Tomography Images

Kenan Morani

Electrical and Electronics Engineering Department, Izmir Democracy University, Izmir, Turkey
[1] kenan.morani@gmail.com

**ABSTRACT-** This paper extends our previous method for COVID-19 diagnosis proposing a more accurate solution for COVID-19 detection from computed tomography images. Here, CT scan slices were processed aiming at localization of the lung areas- the Region Of Interest (ROI). To achieve that, firstly uppermost and lowermost slices of each CT scan were removed keeping 60% of each patient's slices. Secondly, all slices were manually cropped to bring more focus to the lung areas in the images. Next, Slices of CT scans were resized to 224x224 and input into an Xception transfer learning model. Both the architecture and pre-trained weight of the Xception were leveraged, and the output of the model was modified to suit a binary classification task. The modified transfer learning model takes the final diagnostic decisions. The results of the validation CT of the COV19-CT database are promising both at the slices and patient level. Higher validation accuracy was achieved at the slices level and higher validation accuracy and macro F1 score at the patient level compared to our previously proposed solution and compared to other alternatives on the same dataset.

## 1. Introduction

The unprecedented global challenge posed by the COVID-19 pandemic has underscored the critical need for advanced diagnostic methodologies to effectively curb the virus's spread. Among these methodologies, Computed Tomography (CT) imaging has emerged as a vital tool in providing detailed insights into the manifestations of the disease. In this context, the utilization of CT scan images has proven instrumental in detecting the presence of the virus and understanding its impact on the respiratory system. The intricate details captured by CT scans offer a comprehensive view of the pulmonary structures, making them invaluable for early and accurate diagnosis [1].

To address the urgency of timely and precise COVID-19 diagnosis, the integration of advanced computational techniques has become imperative. Deep learning, particularly through the lens of transfer learning, has demonstrated remarkable potential in enhancing diagnostic accuracy and efficiency. Transfer learning, a paradigm that leverages pre-trained models to expedite the learning process, plays a pivotal role in the analysis of medical images. In the realm of COVID-19 diagnosis, these approaches contribute significantly by automating feature extraction and pattern recognition, thereby streamlining the diagnostic workflow [2].

The importance of employing deep learning, and specifically transfer learning, lies in its ability to decipher complex patterns within CT images associated with COVID-19 manifestations. By building on knowledge gained from related tasks, transfer learning models quickly adapt to the unique characteristics of COVID-19 pathology. This not only expedites the diagnostic process but also enhances the accuracy of identifying subtle nuances in CT scan images indicative of viral infection. The potential of these approaches to revolutionize COVID-19 diagnosis underscores the need for continued research and development in this domain [3].

Building upon existing methodologies [4], our proposed solution seeks to further elevate the accuracy of COVID-19 diagnosis through a refined approach. We extend our previous method by introducing advanced image processing techniques applied before inputting data into a modified Xception model. This innovative preprocessing strategy aims to address specific challenges encountered in CT scans, such as non-representative slices, by systematically removing them. Moreover, the proposed method involves manual cropping of images to retain only the pertinent lung areas in each slice, focusing the analysis on regions critical for COVID-19 detection.

The rationale behind this image-processing technique is to enhance the effectiveness of the subsequent transfer learning model. By providing the model with refined and relevant input data, we aim to optimize its performance in discerning COVID-19-related patterns. Preliminary results indicate that this preprocessing significantly contributes to the accurate identification of viral manifestations in CT scans, reinforcing the potential of transfer learning in the context of COVID-19 diagnosis. In this paper, we present a detailed exploration of our proposed solution, including methodological intricacies, experimental outcomes, and the broader implications for the advancement of diagnostic capabilities in the fight against the COVID-19 pandemic.

## 2. Material and Method

The method follows in two parts. Our previous solution did not include the image processing part in the solution. This method aims at adding this part before using a classifier for diagnosing the disease.[1]

**Images Processing.** In the pursuit of refining the input data for our classifier, we implemented two key image processing techniques, each designed to bolster the model's accuracy and minimize misclassifications at the patient level.

Firstly, Selective Slice Removal was applied. Our first image processing technique involves the judicious removal of slices from each CT scan, strategically aimed at preserving only those slices that distinctly represent COVID-19 manifestations. Specifically, we systematically eliminate 40% of the slices in each CT scan, removing an equal number of uppermost and lowermost slices. This curation ensures that the retained slices are central and, therefore, more likely to encapsulate the characteristic features of COVID-19 pathology within the patient. By discarding non-representative slices, we intend to enhance the model's focus on the most relevant sections of the CT scan, thereby contributing to a more accurate and nuanced classification. This selective slice removal process aligns with the overarching goal of tailoring the input data to the unique characteristics of COVID-19 presentations in each patient. Recognizing that the upper and lower extremes of CT scans may not consistently capture the crucial features indicative of the virus, our approach optimizes the dataset to foster a more precise and targeted analysis.

Secondly, Manual Cropping for Lung Area Emphasis was applied. The second facet of our image processing strategy involves manual cropping of all slices, transitioning from the original 512x512 dimensions to a standardized size of 227x300. This deliberate resizing is not merely an arbitrary adjustment; rather, it is a meticulous act aimed at emphasizing the lung areas within each slice. By focusing on the anatomical regions most pertinent to COVID-19 detection, we facilitate the classifier in honing its attention to the key structures indicative of the viral infection. The choice to manually crop each slice aligns with the understanding that the nuances of COVID-19 pathology often manifest prominently in the lung areas. This deliberate act of slice cropping enhances the classifier's ability to discern subtle patterns associated with the virus, ultimately contributing to heightened diagnostic accuracy.

Finally, all slices were resized to 224x224, adding more channels to reach 3 as a standard input image for the Xception model.

**Modified Xception Model Classifier.** In our previous methodology in [4], we used an Xception model with a modified output to make final diagnostic decisions. Fig. 1 Shows the Xception model architecture.

---

[1] https://github.com/IDU-CVLab/COV19D_2nd

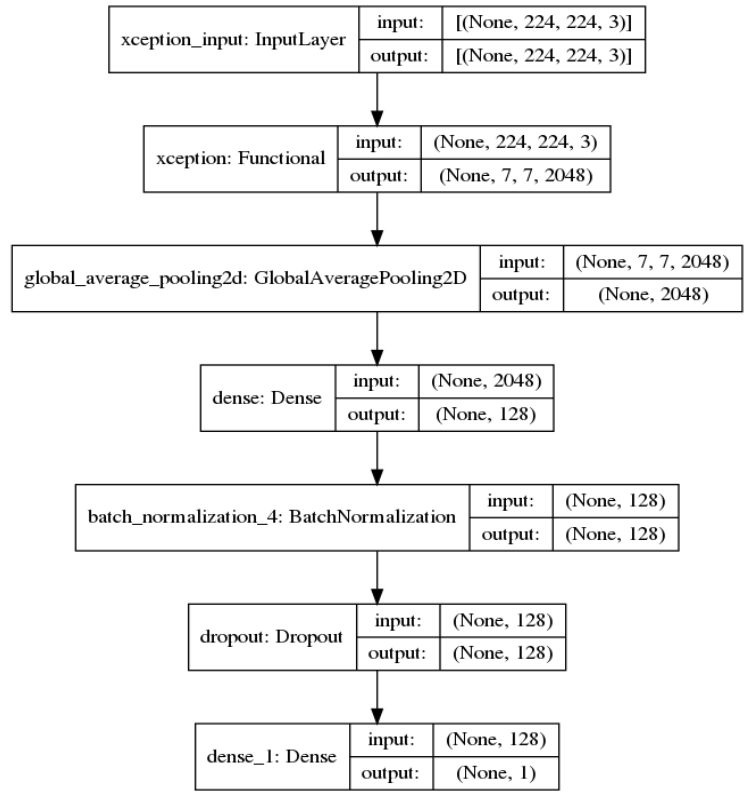

**Fig 1** Xception model architecture

The final layer's output represents the class probability of being a non-COVID-case slice. This class probability is then compared against a predefined threshold, determining the slice's classification as COVID or non-COVID. These individual slice-level determinations collectively lead to patient-level diagnoses, as explained in the latter sections of the paper. Several class probability thresholds were explored to optimize performance, and their effects were evaluated on the validation set of the COV19-CT database.

For model compilation, we employed the Keras platform, utilizing the "Adam" optimizer initialized with a learning rate of 0.001. The loss function was set as "binary cross entropy". Our model was trained across 15 epochs, a determination that emerged from rigorous experimentation. During these trials, it was observed that further increasing the epoch count resulted in only marginal improvements in validation loss over a prolonged timeframe.

Training the CNN model, with a batch size of 128, across the 13 epochs necessitated approximately 7 days of computation. This was facilitated on a workstation operating a GNU/Linux system, equipped with 64GiB of system memory and powered by an Intel(R) Xeon(R) W-2223 CPU @ 3.60GHz processor. These specifications offer insights into the computational resources and time investment involved in achieving our model's refined performance for COVID-19 detection.

**The Dataset.** The dataset utilized in this investigation is an extension of the COV19-CT-DB, playing a crucial role by providing a comprehensive collection of CT scans essential for the detection of COVID-19 [5-6-7-8-9-10]. This dataset encompasses a significant number of CT scans, encompassing 1,650 instances of COVID-19 and 6,100 non-COVID-19 cases. This balanced distribution facilitates a robust assessment of the performance of the proposed method across different classes.

What distinguishes the 'COV19-CT-DB' dataset is not only its size but also its diversity, encompassing variations in the number of cases and the variability in COVID-19 manifestations. The CT scans in the dataset have been meticulously labeled by a panel of experts, each possessing over 20 years of experience, ensuring the accuracy and reliability of the labels, crucial for the construction and evaluation of machine learning models.

The dataset's diversity, spanning a spectrum of COVID-19 and non-COVID-19 cases, introduces unique challenges and opportunities. Given that COVID-19 exhibits a range of manifestations, capturing this variability becomes vital for the development of an effective detection model. The inclusion of cases with varying degrees of lung involvement and diverse clinical presentations in the 'COV19-CT-DB' dataset mirrors the real-world complexity of COVID-19 instances.

With its comprehensive labeling, extensive size, and diversity, the 'COV19-CT-DB' dataset serves as an ideal foundation for assessing the effectiveness of the proposed method. Its suitability arises from its capacity to rigorously evaluate the model's performance on diverse cases, ensuring not only accuracy but also robustness in identifying COVID-19 instances within varying clinical contexts. Each CT scan consists of a variable number of slices, ranging from 50 to 700, and access to this dataset is facilitated through the "ECCV 2022: 2nd COV19D Competition." Table 1 illustrates the distribution of COVID-19 and non-COVID-19 cases for our study.

**Table 1** Distribution of cases in the COV19-CT Database

| Annotation | Training Data | Validation Data |
|---|---|---|
| COVID-19 CT cases | 882 | 255 |
| Non-COVID CT cases | 1110 | 468 |

**Performance Evaluation.** The proposed model was evaluated via the COV19-CT-DB database using accuracy, macro F1 score, and confidence interval.

The accuracy is calculated as in (1).

$$Accuracy = \frac{True\ Positives + True\ Negatives}{True\ Positives + False\ Positives + True\ Negatives + False\ Negatives} \quad (1)$$

Where positive and negative cases refer to COVID and non-COVID cases.

The macro F1 score was calculated after averaging precision and recall matrices as in (2).

$$Macro\ F1 = \frac{2 \times average\ precision \times average\ recall}{average\ precision + average\ recall} \quad (2)$$

Furthermore, to report the confidence intervals of the results obtained, the Binomial proportion confidence intervals are used. The confidence intervals were used to check the range variance of the reported results. The residuals of the interval can be calculated as in (3) [28-29].

$$Radius\ of\ Interval = z \times \sqrt{\frac{value\ considered \times (1 - value\ considered)}{n}} \quad (3)$$

where z is the number of standard deviations from the Gaussian distribution and n is the number of samples.

## 3. Results

The results of our methodology are discussed on the validation set both at the slice level and at the patient level.

**Results at Slices Level.**

Table 2 shows the training performance for different metrics.

**Table 2** Performance results of the training at the slice level

| Performance metric | Score |
|---|---|
| Average training accuracy | 97.30% |
| Average recall | 0.917 |
| Average precision | 0.909 |
| Macro F1 Score | 0.891 |
| Validation Accuracy | 88.48% |

To calculate the confidence interval for the resulting validation accuracy score (0.8848), equation 3 was used. In the equation, z is taken as z=1.96 for a significance level of 95%. By that, we can obtain the confidence interval keeping in mind that the number of samples (slices) in the validation set is 30235, to be approximately 0.0036. With that, the validation accuracy score can be said to be 0.8848 ± 0.0036.

Using the above-mentioned method, predictions were made through different class probability thresholds. These thresholds are compared to the model output. The model has only one output, which is the probability of the slice being a non-COVID slice. After that, the majority voting method for each CT scan was deployed to decide whether the patient belonging to that CT scan was COVID-19 positive or not. Figure 1 shows performance results on the validation set at the patient level for four different class probability thresholds. The comparison was made in terms of the validation accuracy and the macro F1 score.

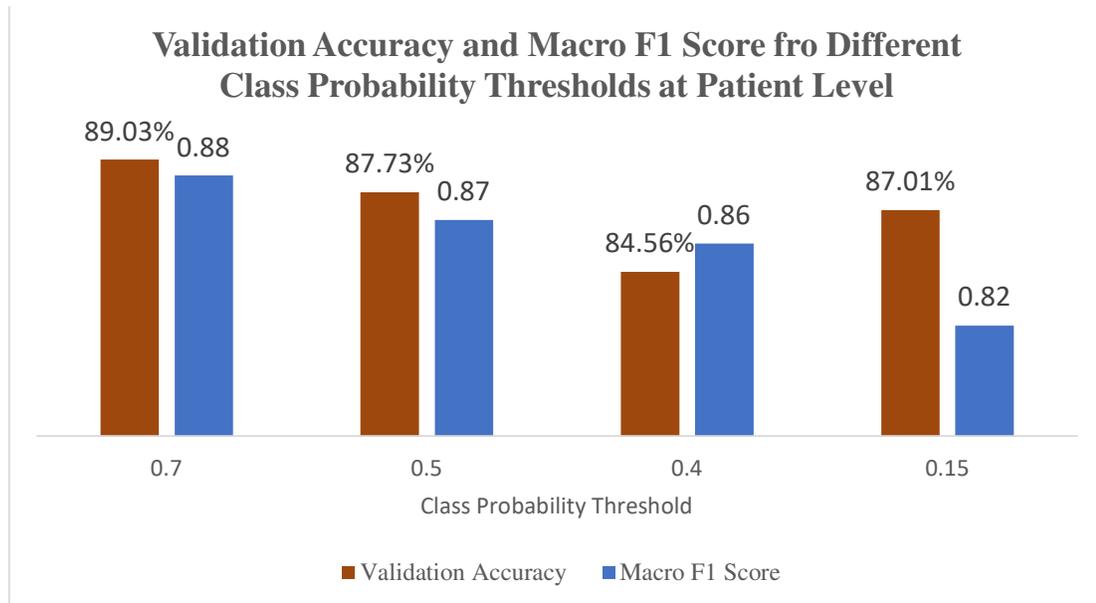

Figure 1 Model performance against different class probability thresholds on the validation set

The findings indicate that, among the three suggested class probability thresholds, the 0.7 threshold level gives the best performance. This holds when considering both validation accuracy and validation macro F1 score. Consequently, our proposed approach exceeds the baseline model score, as reported in [5], in terms of macro F1 score, achieving a score of 0.88 on the validation set. Furthermore, it well exceeds our previous solution without image processing [4].

Further, our macro F1 score is compared to other alternatives on the same dataset in Table 3.

Table 3 Average macro F1 score results from the comparison of validation and test partitions

| The Method | Validation set |
|---|---|
| Cov3d [11] | 0.947 |
| BERT method [12] | 0.916 |
| Base Line [5] | 0.770 |
| **Proposed method (best performance)** | **0.880** |

## 2 Conclusion

In conclusion, we have extended our previous method by adding image processing techniques to CT scan slices before classification. The image processing techniques included uppermost and lowermost slice removal in each CT scan and manual rectangular cropping to the original Slices to focus on the lung areas.

For classification our method uses the same transfer learning approach we introduced in our previous study; a modified Xception model classifier. Proposing the image processing techniques in this paper gave better performance compared to our previous solution, the baseline solution, and many other alternatives on the same dataset.

## Acknowledgment

The authors acknowledge the medical staff who rigorously annotated the dataset used for this study.

## Declarations

**Funding statement.** No funding was provided for this study. No conflict of interest is to be reported.
**Additional information.** The code related to this study can be found on GitHub at https://github.com/IDU-CVLab/COV19D_2nd.

## Data and Software Availability Statements

Data are provided and should be ordered from the corresponding entity as stated in the 'Material and Methodology' section, 'Dataset' subsection.